\documentclass{moriond}

\def\be{\begin{equation}}
\def\ee{\end{equation}}
\def\bea{\begin{eqnarray}}
\def\eea{\end{eqnarray}}

\newcommand{\Photo}{}

\begin{document}
\vspace*{4cm}
\title{Experimental signatures of subtleties in the Brout-Englert-Higgs mechanism}

\author{Axel Maas}

\address{Institute of Physics, NAWI Graz, University of Graz, Universit\"atsplatz 5, 8010 Graz, Austria}

\maketitle\abstracts{Subtle, but long-known, field-theoretical aspects require a more refined treatment of gauge theories involving a Brout-Englert-Higgs effect. This refinement can be done analytically using the Fröhlich-Morchio-Strocchi mechanism. In the standard model, this leads to slight, but in principle detectable, quantitative changes in observables. This can have significant implications for current and future colliders, which are investigated for a few sample processes.}

\section{Introduction}

There exists a long-standing tension between formal aspects of quantum gauge theories and their very successful treatment using perturbation theory, especially for electroweak physics \cite{Maas:2017wzi,Maas:2023emb}: Even in the presence of a Brout-Englert-Higgs effect perturbation theory is on principle grounds ill-defined for multiple reasons. This tension is resolved by the Fr\"ohlich-Morchio-Strocchi (FMS) framework \cite{Maas:2017wzi,Frohlich:1980gj,Frohlich:1981yi}. It replaces the elementary particles of perturbation theory with bound states of the elementary particles and the Higgs in matrix elements. A perturbative treatment is still possible, if the expansion is performed around the bound states. This is the FMS mechanism. Perturbation theory augmented by the FMS mechanism then yields that ordinary perturbation theory is the first term of a finite sum. However, the specific features of the standard model suppress the contribution of the other terms, yielding only very small quantitative deviations. This resolves the tension, agrees with current experimental results, and has been confirmed in full non-perturbative lattice simulations of truncated standard models \cite{Maas:2017wzi,Afferrante:2020fhd}.

But the other terms are not zero \cite{Dudal:2020uwb,Maas:2020kda,Jenny:2022atm,Maas:2022gdb}. Thus, they add unaccounted for standard model background, which needs to be included. Also, beyond the standard model the effects can become qualitative, and dominate over the non-augmented perturbative contribution \cite{Maas:2017wzi,Maas:2023emb,Sondenheimer:2019idq}. This alters potentially fundamentally model building. Hence, just relieving the tension by the FMS framework is not enough. (FMS-mechanism) augmented perturbation theory needs to be used instead of standard perturbation theory.

To establish both the necessity and the adequacy of augmented perturbation theory requires besides lattice support experimental tests. The following provides some first possible tests, though increased precision needs still to be achieved for a final confrontation and eventually confirmation.

\section{Augmented perturbation theory}

Augmenting perturbation theory by the FMS mechanism requires only very few additions. The FMS framework first requires that the asymptotic states are not created by elementary fields, but rather by manifestly gauge-invariant composite fields, carrying global quantum numbers. E.\ g., a physical, left-handed electron is described by the composite operator $L=X^\dagger l$, where $l=(e^L,\nu^L)^T$ is the elementary left-handed doublet lepton field and $X$ a matrix-valued representation of the Higgs field \cite{Frohlich:1980gj,Frohlich:1981yi,Afferrante:2020fhd}. The gauge-invariant $L$ is a doublet of the global (custodial) symmetry carried by the Higgs, one component being, e.\ g., the physical electron and the other the physical electron-neutrino. From these matrix elements for (scattering) processes are build. E.\ g.\ for physical (massless left-polarized) electron annihilation into (massless left-polarized) muons this is $\langle \overline{L}_1^1 L_1^1\overline{L}_1^2 L_1^2\rangle$, where the upper (lower) index is the generation (global Higgs charge) index. These are the same matrix elements as would be obtained in an $n$PI approach \cite{MPS:unpublished}.

For an augmented perturbative treatment the Higgs field in the composite operators is expanded in a suitable gauge in its fluctuations $\eta$  around its vacuum expectation value $v$, e.\ g.\ as $X=v\bf{1}+\eta$. This yields a finite sum of matrix elements, which are individually gauge-dependent, but their sum is gauge-invariant, order-by-order in the perturbative expansion \cite{Maas:2020kda}. E.\ g.\ for the scattering process before this yields
\begin{equation}
\left\langle \overline{L}_1^1 L_1^1\overline{L}_1^2 L_1^2\right\rangle=v^4\left\langle \bar{e}^L e^L \bar{\mu}^L \mu^L\right\rangle+v^3\left\langle\underbrace{(\eta^\dagger l^1)_1}e^L \bar{\mu}^L \mu^L+\textrm{permutations}\right\rangle+\mathcal{O}\left(\frac{s}{v^2}\right),\label{xs}
\end{equation}
\noindent where the underbraces identifies a composite operator and permutations indicate that the composite operator appears for each of the four leptons in turn. It is important to note that this implies that neutrino components of $l$ also appear in the matrix elements, which will be decisive in section \ref{s:high}. The first term is hence the matrix element of perturbation theory, while the other terms of this finite sum are additional terms coming from the FMS mechanism. They are suppressed in powers of $\sqrt{s}/v$, where $s$ is the characteristic energy scale of the process, e.\ g.\ the center-of-mass energy. At this point, this is an exact rewriting. The sum is still manifestly (and non-perturbatively) gauge-invariant, even if this is no longer the case for the individual terms. However, it shows that the elementary fields only are the leading term in $s/v^2$. Thus the elementary valence fields of the composite operator are the FMS dominant constituent, and the Higgs valence field acts at first order in $s/v^2$ entirely as a spectator.

In augmented perturbation theory, the matrix elements in (\ref{xs}) are now expanded as in ordinary perturbation theory, including for the composite operators \cite{MPS:unpublished}. If the full sum is kept, this automatically guarantees that gauge invariance is manifestly maintained order-by-order \cite{Maas:2020kda}. It has been explicitly demonstrated that, e.\ g., the gauge-dependence of residua, which appear in perturbation theory, is canceled by the other terms \cite{Dudal:2020uwb,Maas:2020kda}.

Resummation needs to respect this structure \cite{Dudal:2020uwb,Maas:2020kda}, which can be achieved by using the Dyson-Schwinger equations \cite{MPS:unpublished}. A non-trivial complication is that external wave-functions in the LSZ formalism need to be replaced by bound state amplitudes. These, however, can also be calculated within augmented perturbation theory \cite{MPS:unpublished}.

\section{Signatures}

The results are found to have a structure which is familiar from hadron physics  \cite{Maas:2023emb}. If two such weak gauge-invariant bound states interact, three different regimes are observed. If all relevant energy scales are small compared to the vacuum expectation value, the bound state is probed as a whole. In the intermediate regime, where the relevant energy scales are of order of the vacuum expectation value, the bound state is dominated by the FMS-dominant valence particle. In a sense, it is actually here where the biggest difference is observed to hadrons, as for hadrons all valence particles contribute. However, because only the FMS-dominant constituent contributes, this is actually the kinematic region in which the results of augmented perturbation theory most closely resemble the non-augmented perturbative results. Thus, it is also the regime in which a distinction is most challenging. Finally, at very high energies, the other valence (and eventually sea) constituents start to be probed. Each of the three regimes has its own distinctive features, and it is therefore worthwhile to treat them in turn.

\subsection{Low energies}\label{s:low}

At very low energy, the bound state is probed as a whole. This determines also the response in reactions. Most notably, the extension of the bound state becomes relevant. This yields a characteristic modification of the form factor \cite{Maas:2018ska}. Lattice results suggest a characteristic extension of a few tens of GeV \cite{Jenny:2022atm,Maas:2018ska}, making the bound states about a factor 5 more compact, in relation to their mass, as the proton. Such non-trivial form factors modify the (differential) production rate, and thus also show up if the particle is appearing as an intermediate state. Even for unstable particles, like weak ones, this makes the effect experimentally accessible.

While it is possible to calculate the process in augmented perturbation theory, its higher order nature makes this yet challenging. However, it is possible to access the effect by combining lower-order augmented perturbation theory with lattice results, at least for a truncated standard model \cite{Jenny:2022atm}. In this case, it is possible to study the process $VV\to V^{'}V^{'}$, where the $V^{(')}V^{(')}$ are any combinations of the pairs $ZZ$ and $W^+W^-$. Besides a genuine four-vector-boson interaction, this contains both vector boson fusion and vector boson scattering (VBS) with intermediate Higgs or $Z$.

\begin{figure}
 \includegraphics[width=0.5\textwidth]{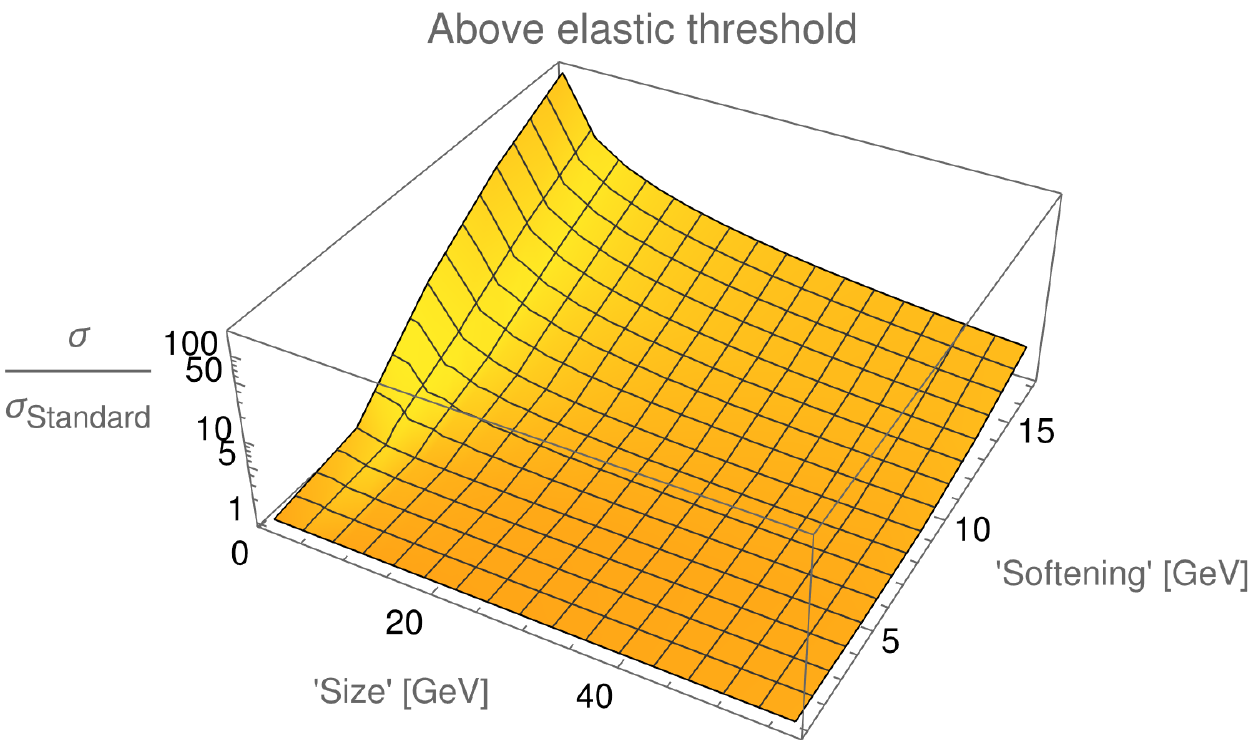}\includegraphics[width=0.5\textwidth]{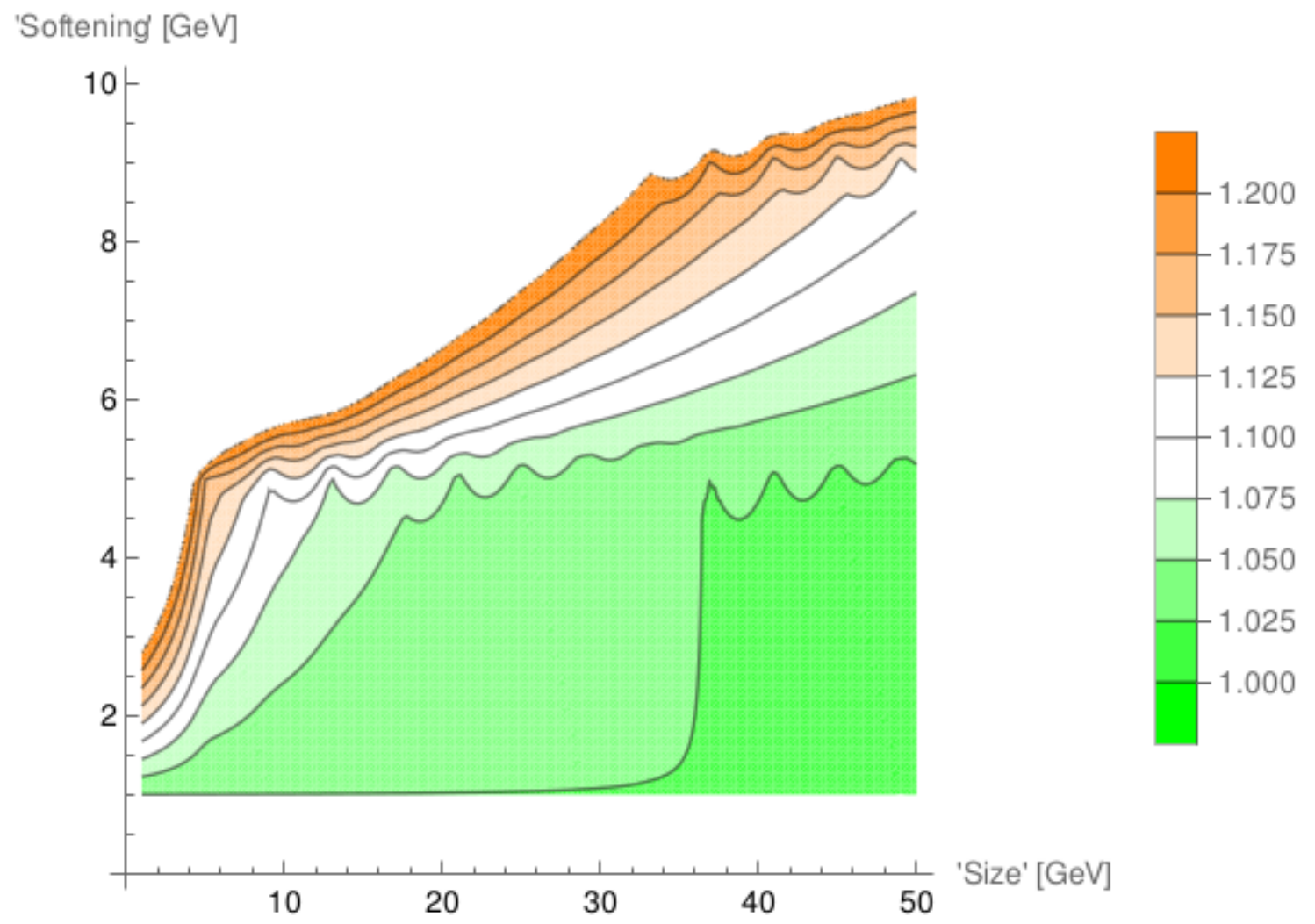}
 \caption{\label{vbs}Left panel: The change of the cross section compared to the Born-level perturbative cross section in the process $VV\to V^{'}V^{'}$ integrated between the elastic threshold and 1.2-times the elastic threshold in the $VV$ system. Size and softening correspond to the inverse scattering length and first-nonvanishing momentum-dependent coefficient in the universal threshold expansion, respectively. Right panel: An exploratory estimate of the exclusion limits for the parameter values, i.\ e.\ which ratio of the cross sections is still compatible with which extension parameters, including estimates of non-electroweak background.}
\end{figure}

Considering the process just above the threshold, i.\ e.\ just above $2m_V$, any extended state below threshold will show up in the corresponding partial wave. The effect will be stronger the closer the mass of the state is to the threshold, and the larger the extent is. The exact values depend, of course, on the theory in question, and also the quantitative value of its coupling constants. In principle, the modification could then be measured, and by direct comparison to the calculations the extent, or even density profile, be determined. A sample of the modification of the cross section is shown in figure \ref{vbs}.

Experimentally, this kinematics is already measured, though as a control region, in the context of the off-shell determination\footnote{Direct measurements of, e.\ g., VBS processes use enrichment techniques on the event sample \cite{ATLAS:2020nlt,CMS:2022woe}. However, even the angular dependencies in this kinematic region change \cite{Jenny:2022atm}, and this may therefore affect the signal.} of the width of the Higgs \cite{ATLAS:2023dnm,CMS:2022ley}. Electroweakly initiated processes make up only about 10\% of the total signal. The data is furthermore only available within one bin from threshold to about 1.2-times the threshold energy at ATLAS \cite{ATLAS:2023dnm} or in three equal bins between about 1.2-times to 1.5-times the threshold energy at CMS \cite{CMS:2022ley}. The experimental uncertainty in these bins is about 10\%, and thus of the same size as the electroweak-initiated processes, and consistent within this uncertainty window with the perturbative expectation. The results in the left panel of the figure \ref{vbs} correspond to the result in the same bin as the ATLAS result \cite{ATLAS:2023dnm}. In the right panel of figure \ref{vbs} the expected difference between augmented perturbation theory and non-augmented perturbation theory is shown, rescaled to the total electroweak signal and standard model parameters. Note that while many approximations are going into this, no assumption on physics are made. This is thus a standard model prediction. It is visible that the size is much less restricted than the softening. Within the statistical errors, even the numerical lattice results of the truncated standard model in lattice calculations remains consistent, 39(10) GeV 'size' and a 'softening' of 12(10) GeV \cite{Jenny:2022atm}. Especially the latter is likely too high as in the lattice calculations \cite{Jenny:2022atm} the Higgs masses were with 150 GeV higher than in the standard model. In the three CMS bins, the expected signal is much smaller \cite{Maas:2023emb}, and thus likely no distinction will be possible without much more precise measurements.

\subsection{Intermediate energies}

If the relevant energy scale of a process is much larger than the 'size', but still not much larger than the vacuum expectation value of the Higgs, an intermediate regime is reached. Here, kinematic suppression \cite{Maas:2017wzi,MPV:unpublished} of the further terms in the FMS sum occurs, while the energies are too large to be able to resolve the composite state effects. As such, the results are essentially given by the one FMS-dominant valence particle.

\begin{figure}
\begin{minipage}[c]{0.6\textwidth}
\includegraphics[width=\textwidth]{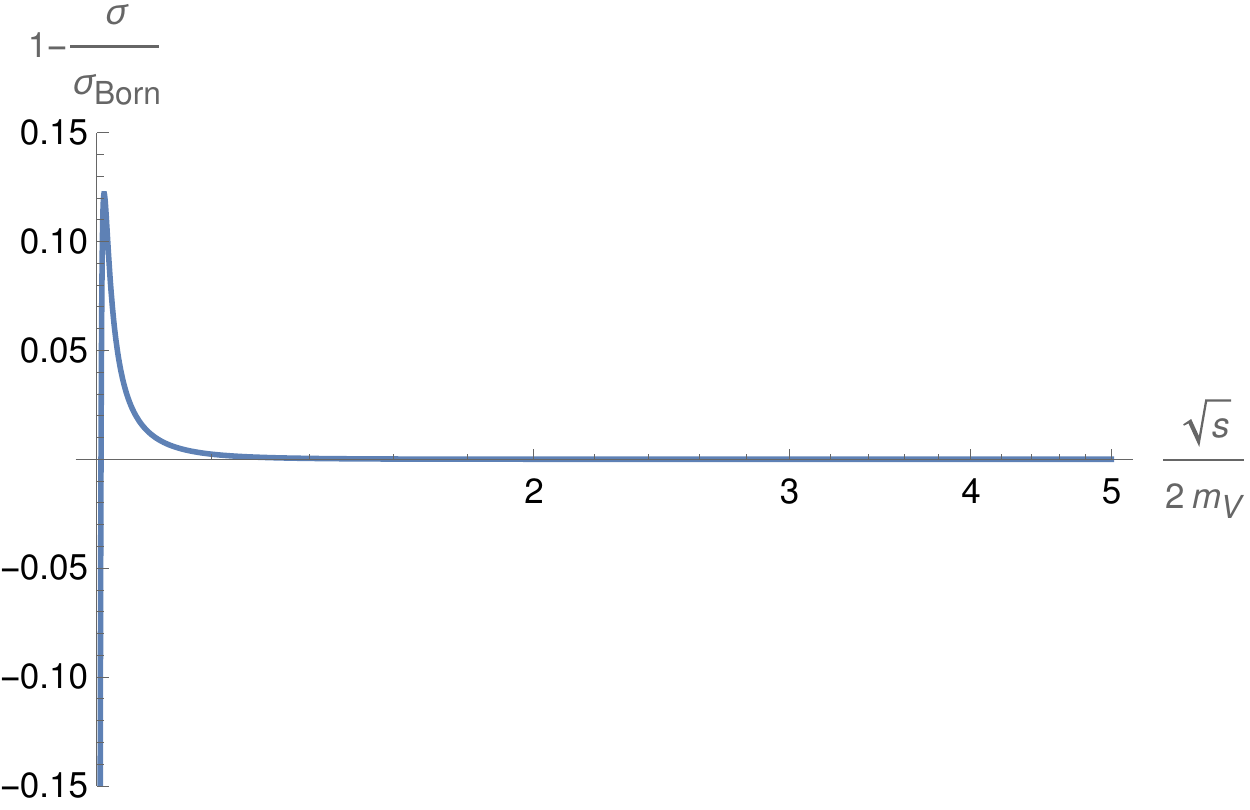}\\
 \end{minipage}\hfill
 \begin{minipage}[c]{0.3\textwidth}
 \caption{The momentum dependence of the deviation of the composite cross section to the elementary Born-level cross section in the truncated standard model for center-of-mass energies in the $VV$-system in units of the threshold energy. The 'size' parameter is 40 GeV and the 'softening' is 5 GeV.}
 \label{int}
 \end{minipage}\hfill
\end{figure}

This has been seen explicitly in calculations on \cite{Jenny:2022atm,Maas:2018ska} and off the lattice \cite{Dudal:2020uwb,Maas:2020kda}. It is visualized for the process discussed in section \ref{s:low} for parameters consistent both with the lattice results \cite{Jenny:2022atm} and the ATLAS data \cite{ATLAS:2023dnm} in figure \ref{int}. It is visible, that already at 20\% more energy than threshold, i.\ e.\ where CMS data is available \cite{CMS:2022ley}, the difference has dropped below 1\%. Thus, the choice of kinematic regions is crucial to detect the effect.

\subsection{High energies}\label{s:high}

If the energies are substantially larger than the vacuum expectation value the remaining valence (Higgs) particles and the electroweak sea particles are also probed, very much like in deep-inelastic scattering \cite{Maas:2023emb}. This probes into the TeV range of parton energies, which only now starts to become accessible at appreciable parton luminosities at the LHC. In this range sizable deviations are again expected, due to the presence of the additional valence particles \cite{Dudal:2020uwb,Maas:2020kda,Maas:2022gdb}. These add to the electroweak sea effects \cite{Bauer:2017isx}.

\begin{figure}
\begin{minipage}[c]{0.6\textwidth}
\includegraphics[width=\textwidth]{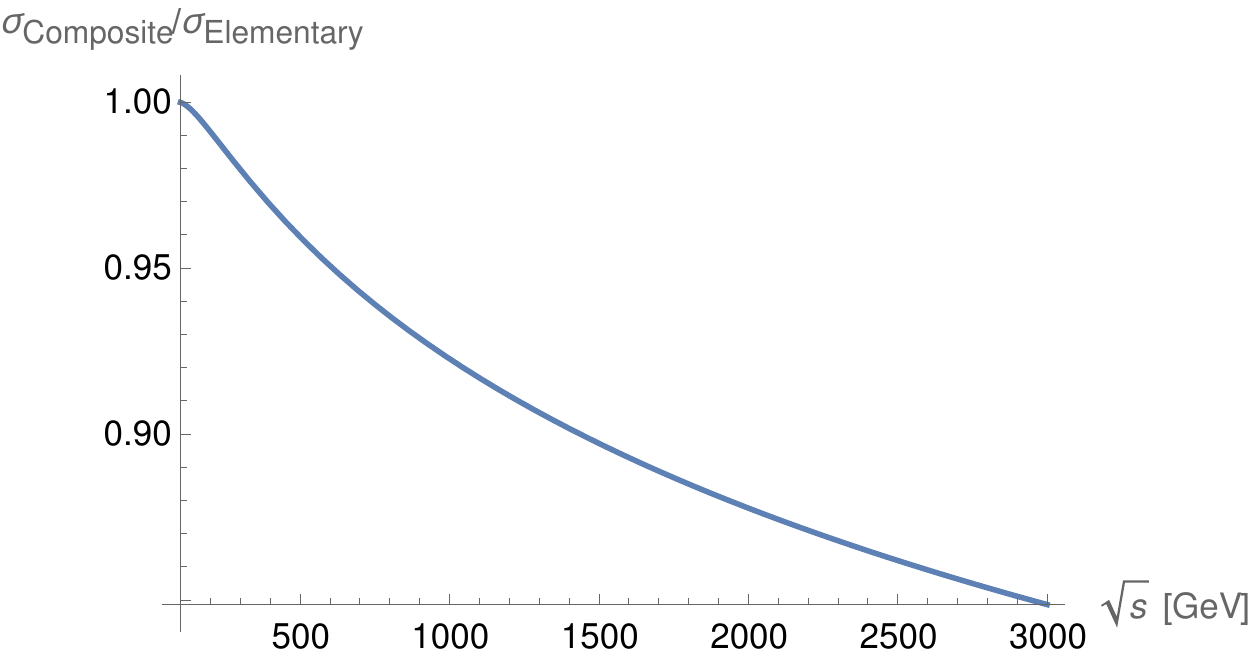}\\
 \end{minipage}\hfill
 \begin{minipage}[c]{0.3\textwidth}
 \caption{Sketch of the impact of restoring the Bloch-Nordsieck theorem ($\sigma_\textrm{Composite}$) compared to the non-augmented perturbative result ($\sigma_\textrm{Elementary}$) in the process $e^-e^+\to$ hadrons as a function of center-of-mass energy, including strong corrections.}
 \label{bn}
 \end{minipage}\hfill
\end{figure}

The most prolific consequence is likely to be seen in collinear physics. At such high energies the vector bosons and the Higgs are effectively massless. They can thus be radiated similar to photons in both the initial state and final state. Because perturbation theory starts form elementary states, Bloch-Nordsieck cancellation of, effectively, infrared divergences due to real and virtual emissions of vector bosons cannot happen \cite{Ciafaloni:2000df}. This leads to a change of the cross section for the interaction of left-handed fermions of the same order as the strong corrections. In the manifest gauge-invariant FMS framework the cancellations happen, as here the same cancellation mechanism as in the electromagnetic case takes effect \cite{Maas:2022gdb}, as the composite operators contain full weak multiplets. E.\ g., for $2l\to 2f$ processes, the cross sections will deviate by a term of order $\ln^2 s/m_W^2$ between the physical, composite scattering, and the perturbative elementary one \cite{Maas:2022gdb,Ciafaloni:2000df}. At future lepton colliders at sufficiently high energies, this would be a marked effect. This is sketched for the case of $e^+e^-\to$ hadrons in figure \ref{bn}. Of course, replacing the initial states by muons for a muon collider would not alter the effect perceptible. Thus, the total cross section of this process drops markedly in comparison to expectations from non-augmented perturbation theory.

Unfortunately, such lepton colliders will not be available on short notice. At the LHC, the same effect is reduced due to the bound state structure of the protons \cite{Maas:2022gdb,Ciafaloni:2000df}. Furthermore, making the proton an electroweak singlet affects also the structure of the parton distribution functions \cite{Maas:2022gdb}. This happens in addition to the non-augmented perturbative electroweak effects from the sea \cite{Bauer:2017isx}. Moreover, this will also affect final state production of electroweak particles \cite{Maas:2022gdb,Platzer:2022nfu}, and would thus need to be taken duly into account. A quantitative prediction would, however, require to take these effects into account already when determining the parton distribution functions \cite{Maas:2022gdb,Fernbach:2020tpa}.

\section{Conclusion}

Taking manifest gauge-invariance into account leads to small, but in principle detectable, deviations from perturbative expansion. Augmenting perturbation theory with the FMS mechanism provides a feasible path to resolve the contributions, making it unnecessary to refer to genuine non-perturbative methods. With more and more tools (becoming) available for augmented perturbation theory \cite{Dudal:2020uwb,Maas:2020kda,Maas:2022gdb,MPS:unpublished,MPV:unpublished}, it will be possible to provide sufficiently precise signatures and predictions for the effects. Since these effects are entirely from the underlying field theory of the standard model, there is no freedom left in these calculations. Hence, failure to detect the effects will invalidate the underlying formal field theoretical understanding of the standard model. Though, of course, it could also be physics beyond the standard model, and both possibilities would then be needed to be disentangled.

However, for physics beyond the standard model, up to and including quantum gravity \cite{Maas:2023emb}, the effects may be much more drastic than for the standard model \cite{Maas:2017wzi,Maas:2023emb}. Even qualitative deviations such as altered spectra would be possible \cite{Maas:2017wzi,Sondenheimer:2019idq}. Thus, any adequate description of physics beyond the standard model need to take into account these effects. Which again seems to be possible using augmented perturbation theory \cite{Maas:2017wzi}.

Ultimately, there is an exciting, guaranteed discovery awaiting. Either, physics beyond the standard model shows up. Or, the predicted effects cannot be found. This would invalidate our formal understanding of quantum gauge theories, requiring a fundamental reworking of the theoretical foundations of particle physics. Or, it is confirmed within the standard model. This would then imply that all particles within the standard model, except for possible right-handed neutrinos, are necessarily composite, extended objects. This possibility leads down to very interesting consequences, both for future model building and conceptual considerations \cite{Maas:2023emb}.

\bibliographystyle{unsrt}    
\bibliography{bib.bib}

\begin{thebibliography}{10}

\bibitem{Maas:2017wzi}
Axel Maas.
\newblock {Brout-Englert-Higgs physics: From foundations to phenomenology}.
\newblock {\em Progress in Particle and Nuclear Physics}, 106:132--209, 2019.

\bibitem{Maas:2023emb}
Axel Maas.
\newblock {\em Rigorous Trails Across Quantum and Classical Physics: A Volume
  in Tribute to Giovanni Morchio}, chapter {The Fr\"ohlich-Morchio-Strocchi
  mechanism: A underestimated legacy}.
\newblock Springer, 5 2023.

\bibitem{Frohlich:1980gj}
J.~Fr\"ohlich, G.~Morchio, and F.~Strocchi.
\newblock {Higgs phenomenon without a symmetry breaking order parameter}.
\newblock {\em Phys.Lett.}, B97:249, 1980.

\bibitem{Frohlich:1981yi}
J.~Fr\"ohlich, G.~Morchio, and F.~Strocchi.
\newblock {Higgs phenomenon without a symmetry breaking order parameter}.
\newblock {\em Nucl.Phys.}, B190:553--582, 1981.

\bibitem{Afferrante:2020fhd}
Vincenzo Afferrante, Axel Maas, Ren\'e Sondenheimer, and Pascal T\"orek.
\newblock {Testing the mechanism of lepton compositness}.
\newblock {\em SciPost Phys.}, 10:062, 2021.

\bibitem{Dudal:2020uwb}
D.~Dudal, D.M. van Egmond, M.S. Guimaraes, L.F. Palhares, G.~Peruzzo, and S.P.
  Sorella.
\newblock {Spectral properties of local BRST invariant composite operators in
  the $SU(2)$ Yang--Mills--Higgs model}.
\newblock {\em Eur. Phys. J. C}, 81(3):222, 8 2020.

\bibitem{Maas:2020kda}
Axel Maas and Ren\'e Sondenheimer.
\newblock {Gauge-invariant description of the Higgs resonance and its
  phenomenological implications}.
\newblock {\em Phys. Rev. D}, 102:113001, 9 2020.

\bibitem{Jenny:2022atm}
Patrick Jenny, Axel Maas, and Bernd Riederer.
\newblock {Vector boson scattering from the lattice}.
\newblock {\em Phys. Rev. D}, 105:114513, 4 2022.

\bibitem{Maas:2022gdb}
Axel Maas and Franziska Reiner.
\newblock {Restoring the Bloch-Nordsieck theorem in the electroweak sector of
  the standard model}.
\newblock 12 2022.

\bibitem{Sondenheimer:2019idq}
Ren\'e Sondenheimer.
\newblock Analytical relations for the bound state spectrum of gauge theories
  with a brout-englert-higgs mechanism.
\newblock {\em Phys. Rev. D}, 101(5):056006, 2020.

\bibitem{MPS:unpublished}
Axel Maas, Simon Pl\"atzer, and Ren\'e Sondenheimer.
\newblock 2023.

\bibitem{Maas:2018ska}
Axel Maas, Sebastian Raubitzek, and Pascal T\"orek.
\newblock {Exploratory study of the off-shell properties of the weak vector
  bosons}.
\newblock {\em Phys. Rev.}, D99(7):074509, 2019.

\bibitem{ATLAS:2020nlt}
Georges Aad et~al.
\newblock {Observation of electroweak production of two jets and a Z-boson
  pair}.
\newblock {\em Nature Phys.}, 19(2):237--253, 2023.

\bibitem{CMS:2022woe}
{Observation of electroweak W$^+$W$^-$ pair production in association with two
  jets in proton-proton collisions at $\sqrt{s}$ = 13 TeV}.
\newblock {\em Phys. Lett. B}, 841:137495, 5 2022.

\bibitem{ATLAS:2023dnm}
Georges Aad et~al.
\newblock {Evidence of off-shell Higgs boson production from $ZZ$ leptonic
  decay channels and constraints on its total width with the ATLAS detector}.
\newblock 4 2023.

\bibitem{CMS:2022ley}
Armen Tumasyan et~al.
\newblock {Measurement of the Higgs boson width and evidence of its off-shell
  contributions to ZZ production}.
\newblock {\em Nature Phys.}, 18(11):1329--1334, 2022.

\bibitem{MPV:unpublished}
Axel Maas, Simon Pl\"atzer, and Fabian Veider.
\newblock 2023.

\bibitem{Bauer:2017isx}
Christian~W. Bauer, Nicolas Ferland, and Bryan~R. Webber.
\newblock {Standard Model Parton Distributions at Very High Energies}.
\newblock {\em JHEP}, 08:036, 2017.

\bibitem{Ciafaloni:2000df}
Marcello Ciafaloni, Paolo Ciafaloni, and Denis Comelli.
\newblock {Bloch-Nordsieck violating electroweak corrections to inclusive TeV
  scale hard processes}.
\newblock {\em Phys. Rev. Lett.}, 84:4810--4813, 2000.

\bibitem{Platzer:2022nfu}
Simon Pl\"atzer and Malin Sjodahl.
\newblock {Amplitude Factorization in the Electroweak Standard Model}.
\newblock 4 2022.

\bibitem{Fernbach:2020tpa}
Simon Fernbach, Lukas Lechner, Axel Maas, Simon Pl\"atzer, and Robert
  Sch\"ofbeck.
\newblock Constraining the higgs valence contribution in the proton.
\newblock {\em Phys. Rev. D}, 101(11):114018, 2020.

\end{thebibliography}

\end{document}